\documentclass[a4paper,11pt]{article}

\usepackage{jheppub}
\usepackage[T1]{fontenc}
\usepackage{hyperref}
\usepackage{graphicx}
\usepackage{amsmath}
\usepackage{epsfig}
\usepackage{subfig}
\usepackage{xcolor}
\usepackage{natbib}

\def\opn{\operatorname}

\title{Comprehensive studies of $\Upsilon$ inclusive production in $Z$ boson decay}

\author[a]{Zhan Sun}
\author{and}
\author[b]{Hong Fei Zhang}
\affiliation[a]{Department of Physics, Guizhou Minzu University, Guiyang 550025, People's Republic of China.}
\affiliation[b]{College of Big Data Statistics, Guizhou University of Finance and Economics, Guiyang 550025, People's Republic of China.}
\emailAdd{zhansun@cqu.edu.cn}
\emailAdd{shckm2686@163.com}

\abstract{In this paper, we present a comprehensive study of $\Upsilon$ inclusive production in $Z$ boson decay, including the first complete next-to-leading-order calculations of the color-octet (CO) contributions. With the inclusion of the newly-calculated remarkable QCD corrections, the CO processes exhibit crucially phenomenological influence on the existing predictions built on the color-singlet mechanism. We also include the exhaustive evaluations of the feed-down contributions, which remained ignored in the literature, and find them to be considerable. Summing up all the contributions, the $\mathcal{B}_{Z \to \Upsilon(nS)+X}$ still notably undershoot the data released by the L3 Collaboration. 
}


\begin{document}

\maketitle

\bibliographystyle{JHEP}

\section{Introduction}\label{intro}

Heavy-quarkonium production in $Z$ boson decay, which has attracted much attention on both theoretical and experimental sides in the past decades \cite{z decay 1,z decay 2,z decay 3,z decay 4,z decay 5,z decay 6,z decay 7,z decay 8,z decay 9,z decay 10,z decay 11,z decay 12,z decay 13,z decay 14,z decay 15,z decay 16,z decay 17,z decay 18,z decay 19,z decay 20,z decay 21,z decay 22,z decay 23,z decay 24,z decay 25,z decay 26,z decay 27,z decay 28,z decay 29,z decay 30,z decay 31,z decay 32,z decay 33}, can offer independent test of the quarkonium-production mechanism and provide references to distinguish different models. A well-known example is that the measurements of $J/\psi$ inclusive production in $Z$ decay apparently exceed the predictions given by the color-singlet (CS) mechanism but coincide with that built on the non-relativistic QCD (NRQCD) framework \cite{z decay 12,z decay 13,z decay 14,NRQCD}, which is regarded as a solid evidence to favor the NRQCD factorization.

In addition to inclusive $J/\psi$ production, L3 Collaboration also measured the total decay width of $Z \to \Upsilon(nS)+X$ \cite{z decay 1},
\begin{eqnarray}
&&\mathcal{B}_{Z \to \Upsilon(1S)+X} < 4.4 \times 10^{-5}, \nonumber \\
&&\mathcal{B}_{Z \to \Upsilon(2S)+X} < 13.9 \times 10^{-5}, \nonumber \\
&&\mathcal{B}_{Z \to \Upsilon(3S)+X} < 9.4 \times 10^{-5}, \nonumber \\
&&\mathcal{B}_{Z \to \Upsilon(1S+2S+3S)+X} = (1.0 \pm 0.4 \pm 0.22) \times 10^{-4}.
\label{eq:data}
\end{eqnarray} 
The leading-order (LO) calculations of the CS process $Z \to b\bar{b}[^3S_1^{[1]}]+b+\bar{b}$ give a prediction $\mathcal{B}_{Z \to \Upsilon(1S+2S+3S)+X} \sim 10^{-6}$ \cite{z decay 2,z decay 3}. Subsequently, Li $et$ $al$. \cite{z decay 4} evaluated the next-to-leading-order (NLO) QCD corrections to this process, and pointed out that the higher-order terms in $\alpha_s$ could give rise to a $24-100\%$ enhancement, depending on the choice of the renormalization scale ($\mu_r$). Recently, we accomplished the NLO calculations of $Z \to b\bar{b}[^3S_1^{[1]}]+g+g$, which is of the same order in $\alpha_s$ as $Z \to b\bar{b}[^3S_1^{[1]}]+b+\bar{b}$, discovering that this process is indispensable \cite{z decay 33}. Summing over all the above contributions, the CS prediction of $\mathcal{B}_{Z \to \Upsilon(1S+2S+3S)+X}$ is just about $10^{-5}$ and therefore incompatible with the L3 measurements.

In order to fill the huge gap between theory and experiment, it is urgent to evaluate the CO contributions, among which only the result for the $\Upsilon(^3S_1^{[8]})$ production via gluon fragmentation\footnote{Actually, the process $Z \to q+\bar{q}+g^{*};g^{*} \to b\bar{b}[^3S_1^{[8]}]$ is only one part of the gluon-fragmentation processes; the other one of the same order in $\alpha_s$ is the loop-induced process $Z \to g+g^{*};g^{*}\to b\bar{b}[^3S_1^{[8]}]$, which was uncalculated in the literature and which will be computed in our present NLO calculations.}, $Z \to q+\bar{q}+g^{*}$ with $g^{*} \to b\bar{b}[^3S_1^{[8]}]$, has been given \cite{z decay 13}. The author of Reference~\cite{z decay 13} claimed that the decay width of $Z\to\Upsilon+X$ can be enhanced by several times with the inclusion of the above gluon-fragmentation contribution alone, which, however, can be attributed to the employment of a very large value of $\langle \mathcal{O}^{\Upsilon}(^3S_1^{[8]})\rangle$. In some recent references~\cite{LDMEs 1, LDMEs 2, LDMEs 3}, this value has been renewed, and found to be more than one order of magnitude smaller than that used in Reference~\cite{z decay 13}. Further, we will find in this paper that the gluon-fragmentation part actually does not dominate the whole CO contributions, as a result, a complete calculation of the CO processes is necessary. To make a comprehensive study, the feed-down contributions from higher excited states, namely $\Upsilon(2S,3S)$ and $\chi_b(1P, 2P, 3P)$, need also to be counted. As a matter of fact, this part is nonnegligible. Taken together, in this paper we will revisit the $\Upsilon$ inclusive production in $Z$ decay by including the first complete NLO studies of the CO processes and the exhaustive evaluations of the feed-down contributions. 

Note that, in comparison with $J/\psi$, $\Upsilon$ may be even more suitable for the investigations of heavy-quarkonium inclusive production in $Z$ decay. The large $b$-quark mass will make the $\Upsilon$ decay products more energetic and thereby more easily detectable. What is more important is, in $Z \to J/\psi+X$, the significant (even dominant) $b$-hadron feed-down contributions will impose great obstacles to the extraction of prompt $J/\psi$ production rates; that is to say, the value of $\mathcal{B}_{b \to J/\psi+X}$ severely affects the precision of fitting $\mathcal{B}_{Z \to J/\psi_{\opn{prompt}}+X}$ \cite{z decay 1}. However, there is no such $b$-hadron feed down contributing in $Z \to \Upsilon+X$, which would be rather beneficial to achieve a precise measurement of $\mathcal{B}_{Z \to \Upsilon_{\opn{prompt}}+X}$. On the theoretical side, the large mass of $b$ quark generally results in a better convergent perturbative series over the expansion of $\alpha_s$ and $v^2$. Considering the large uncertainty of $\mathcal{B}_{Z \to \Upsilon(1S+2S+3S)+X}$ and the lack of definitely measured $\mathcal{B}_{Z \to \Upsilon(nS)+X}$ as listed in equation (\ref{eq:data}), which are primarily attributed to the low $\Upsilon$ production rates at LEP, it is crucial to reperform the measurements with better precision at colliders equipped with much higher luminosities, such as LHC or some planned $Z$ factories. Our state-of-the-art predictions would pave the way for comparisons with the future measurements.

The rest of the paper is organized as follows: Section \ref{cal} is an outline of the calculation formalism. Then, the phenomenological results and discussions are presented in Section \ref{results}. Section \ref{sum} is a concluding remark.

\section{Calculation formalism}\label{cal}

In the context of NRQCD factorization \cite{NRQCD,Petrelli:1997ge}, the decay width of $Z \to \Upsilon(\chi_b)+X$\footnote{$Z \to \chi_b+X$ should be also calculated so as to obtain the $\chi_b$ feed-down contributions.} can be factorized as
\begin{eqnarray}
\Gamma=\hat{\Gamma}_{Z \to b\bar{b}[n]+X}\langle \mathcal O ^{\Upsilon(\chi_b)}(n)\rangle,
\end{eqnarray}
where $\hat{\Gamma}_{Z \to b\bar{b}[n]+X}$ is the perturbative calculable short distance coefficients (SDCs), representing the production of a configuration of the $b\bar{b}[n]$ intermediate state. For $\Upsilon$ production, by neglecting the $\mathcal{O}(v^{4})$ terms, $n$ runs over $^3S_1^{[1]}$, $^1S_0^{[8]}$, $^3S_1^{[8]}$, and $^3P_J^{[8]}$; for the $\chi_b$ case, up to $\mathcal{O}(v^{2})$, $n$ can take $^3P_J^{[1]}$ and $^3S_1^{[8]}$. The universal nonperturbative long distance matrix elements (LDMEs) $\langle \mathcal O ^{\Upsilon(\chi_b)}(n)\rangle$ stand for the probability of $b\bar{b}[n]$ into $\Upsilon(\chi_b)$. 
\subsection{CS framework}
For $n=^3S_1^{[1]}$, there are two LO processes in $\alpha_s$,
\begin{eqnarray}
&&Z \to b\bar{b}[^3S_1^{[1]}]+b+\bar{b}, \nonumber \\
&&Z \to b\bar{b}[^3S_1^{[1]}]+g+g.
\label{cs processes}
\end{eqnarray}
NLO QCD corrections to the first one have been carried out by Li $et~al$. \cite{z decay 4}, and thus we straightforwardly employ their results. Regarding the gluon-radiation process in equation (\ref{cs processes}), i.e., $Z \to b\bar{b}[^3S_1^{[1]}]+g+g$, we reperform the NLO calculations by taking the presently used parameters, based on the formalism described in our recent paper \cite{z decay 33}.

For $n=^3P_J^{[1]}$, the LO processes follow as
\begin{eqnarray}
&&Z \to b\bar{b}[^3P_J^{[1]}]+b+\bar{b}, \nonumber \\
&&Z \to b\bar{b}[^3P_J^{[1]}]+g+g.
\label{csp processes}
\end{eqnarray}
From Reference \cite{z decay 31} we learn that the $\chi_b$ feed-down contributions to $\Upsilon$ production through the two CS processes in equation (\ref{csp processes}) are far less important than that via the CO channel $Z \to b\bar{b}[^3S_1^{[8]}]+X$. In addition, the NLO calculation of $Z \to b\bar{b}[^3P_J^{[1]}]+g+g$ would involve two-loop QCD corrections to $\langle \mathcal O ^{^3P_{J}^{[1]}}(^3S_{1}^{[8]})\rangle$\footnote{The tree-level process $Z \to b\bar{b}[^3P_J^{[1]}]+g+g$ by itself has contained soft singularities, which arise from the attachment of a soft gluon to the P-wave quarkonium and which can only be canceled by introducing the renormalization of $\langle \mathcal O ^{^3P_{J}^{[1]}}(^3S_{1}^{[8]})\rangle$; therefore, two-loop QCD corrections to $\langle \mathcal O ^{^3P_{J}^{[1]}}(^3S_{1}^{[8]})\rangle$ need to be involved in the NLO calculation of $Z \to b\bar{b}[^3P_J^{[1]}]+g+g$.}, which are beyond the scope of this article. Combining the two points, we compute the two $^3P_J^{[1]}$ processes in equation (\ref{csp processes}) only at the tree-level accuracy, which would not contaminate the precision of our NRQCD predictions of $\Gamma_{Z \to \Upsilon+X}$.
\subsection{CO framework} 
\begin{figure}[!h]
\begin{center}
\hspace{0cm}\includegraphics[width=0.95\textwidth]{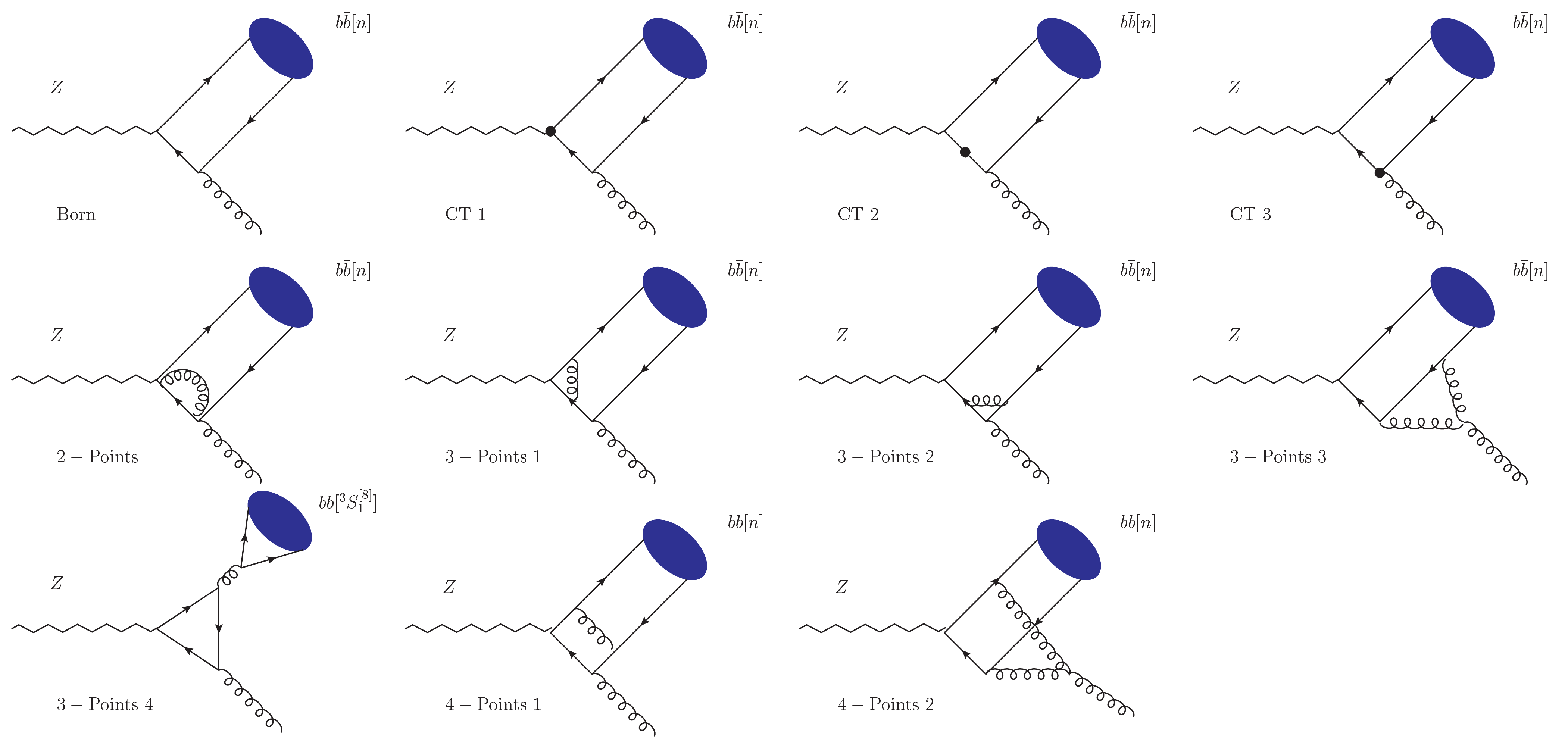}
\caption{\label{fig:virtual}
Typical diagrams for the virtual corrections to $Z \to b\bar{b}[n]+g$. $n=^1S_{0}^{[8]},^3S_{1}^{[8]}$, and $^3P_{J}^{[8]}$.}
\end{center}
\end{figure}

\begin{figure}[!h]
\begin{center}
\hspace{0cm}\includegraphics[width=0.95\textwidth]{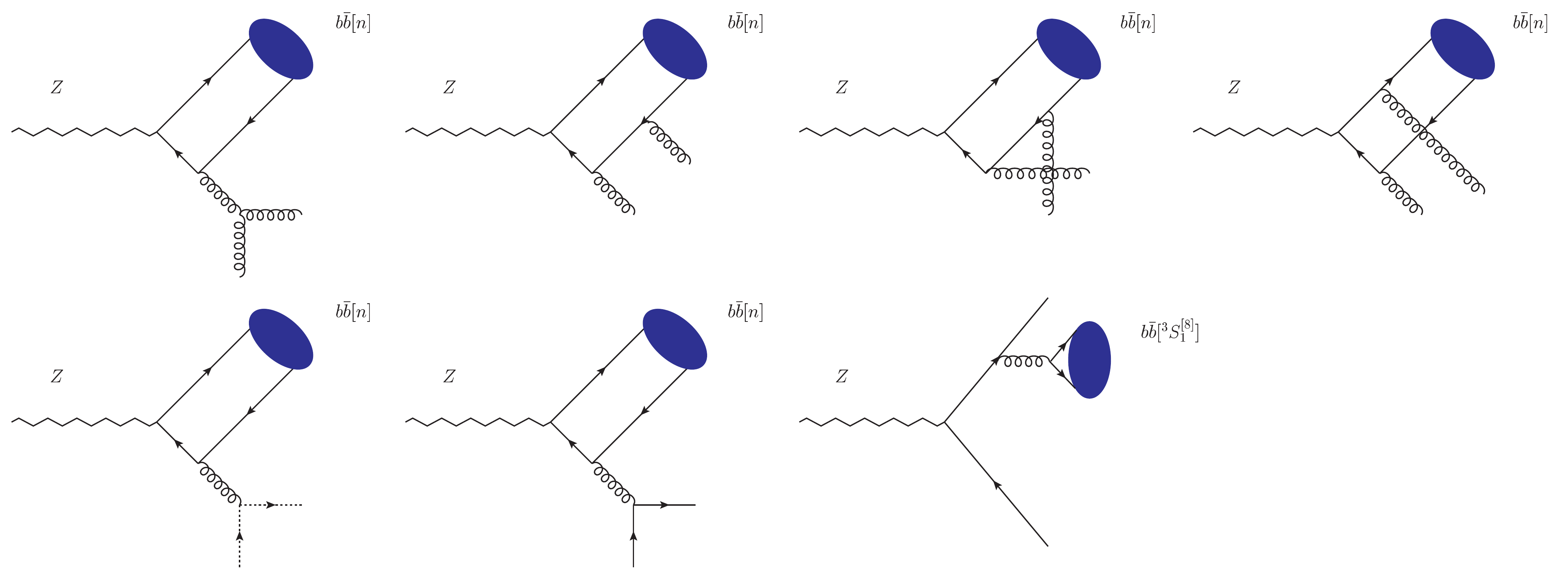}
\caption{\label{fig:real}
Typical diagrams for the real corrections to $Z \to b\bar{b}[n]+g$. $n=^1S_{0}^{[8]},^3S_{1}^{[8]}$, and $^3P_{J}^{[8]}$.}
\end{center}
\end{figure}

In the case of $n=$ $^1S_0^{[8]}$, $^3S_1^{[8]}$, and $^3P_J^{[8]}$, the LO process reads
\begin{eqnarray}
&Z& \to b\bar{b}[n]+g,
\label{CO LO process}
\end{eqnarray}
and the SDCs up to the NLO order in $\alpha_s$ can be written as\footnote{For $^3P_{J}^{[8]}$, the renormalization of $\langle \mathcal O ^{^3P_{J}^{[8]}}(^3S_{1}^{[8]})\rangle$ should be additionally added to eliminate the residually soft singularities.}
\begin{eqnarray}
\hat{\Gamma}=\hat{\Gamma}_{\textrm{Born}}+\hat{\Gamma}_{\textrm{Virtual}}+\hat{\Gamma}_{\textrm{Real}}+\hat{\Gamma}_{\textrm{NLO}^{*}}+\mathcal{O}(\alpha\alpha_s^3).
\end{eqnarray}
$\hat{\Gamma}_{\textrm{Born}}$ refers to the LO processes; $\hat{\Gamma}_{\textrm{Virtual}}$ and $\hat{\Gamma}_{\textrm{Real}}$ represent the virtual and real corrections, respectively. $\hat{\Gamma}_{\textrm{NLO}^{*}}$ denotes the heavy quark-antiquark pair associated processes, i.e., $Z \to b\bar{b}[^1S_0^{[8]},^3S_1^{[8]},^3P_J^{[8]}]+Q+\bar{Q}$ ($Q=c,b$), which are free of divergences and thereby can be directly computed using the standard Monte-Carlo integration techniques. In the following, we just briefly describe the formalism for calculating the virtual and real corrections.

\subsubsection{Virtual corrections}
The virtual corrections consist of the one-loop contributions ($\hat{\Gamma}_{\textrm{Loop}}$) and the counterterm contributions ($\hat{\Gamma}_{\textrm{CT}}$), whose typical diagrams are shown in figure \ref{fig:virtual}; $\hat{\Gamma}_{\textrm{Virtual}}$ can then be expressed as
\begin{eqnarray}
\hat{\Gamma}_{\textrm{Virtual}}&=&\hat{\Gamma}_{\textrm{Loop}}+\hat{\Gamma}_{\textrm{CT}}.
\end{eqnarray}
We utilize the dimensional regularization with $D=4-2\epsilon$ to isolate the ultraviolet (UV) and infrared (IR) divergences. The on-mass-shell (OS) scheme is employed to set the renormalization constants for the heavy quark mass ($Z_m$), heavy quark filed ($Z_2$), and gluon filed ($Z_3$). The modified minimal-subtraction ($\overline{MS}$) scheme is used for the QCD gauge coupling ($Z_g$). The renormalization constants are taken as
\begin{eqnarray}
\delta Z_{m}^{OS}&=& -3 C_{F} \frac{\alpha_s N_{\epsilon}}{4\pi}\left[\frac{1}{\epsilon_{\textrm{UV}}}-\gamma_{E}+\opn{ln}\frac{4 \pi \mu_r^2}{m_b^2}+\frac{4}{3}\right], \nonumber \\
\delta Z_{2}^{OS}&=& - C_{F} \frac{\alpha_s N_{\epsilon}}{4\pi}\left[\frac{1}{\epsilon_{\textrm{UV}}}+\frac{2}{\epsilon_{\textrm{IR}}}-3 \gamma_{E}+3 \opn{ln}\frac{4 \pi \mu_r^2}{m_b^2}+4\right], \nonumber \\
\delta Z_{3}^{OS}&=& \frac{\alpha_s N_{\epsilon}}{4\pi}\left[(\beta_{0}^{'}-2 C_{A})(\frac{1}{\epsilon_{\textrm{UV}}}-\frac{1}{\epsilon_{\textrm{IR}}})-\frac{4}{3}T_F(\frac{1}{\epsilon_{\textrm{UV}}}-\gamma_E+\opn{ln}\frac{4\pi\mu_r^2}{m_c^2}) \right. \nonumber\\
&& \left. -\frac{4}{3}T_F(\frac{1}{\epsilon_{\textrm{UV}}}-\gamma_E+\opn{ln}\frac{4\pi\mu_r^2}{m_b^2})\right], \nonumber \\
\delta Z_{g}^{\overline{MS}}&=& -\frac{\beta_{0}}{2}\frac{\alpha_s N_{\epsilon}}{4\pi}\left[\frac{1} {\epsilon_{\textrm{UV}}}-\gamma_{E}+\opn{ln}(4\pi)\right], \label{CT}
\end{eqnarray}
where $\gamma_E$ is the Euler's constant, $N_{\epsilon}= \Gamma(1-\epsilon) /({4\pi\mu_r^2}/{(4m_b^2)})^{\epsilon}$ is an overall factor in our calculation, $\beta_{0}=\frac{11}{3}C_A-\frac{4}{3}T_Fn_f$ is the one-loop coefficient of the $\beta$ function, and $\beta_{0}^{'}=\frac{11}{3}C_A-\frac{4}{3}T_Fn_{fl}$. $n_f(=5)$ and $n_{fl}(=n_f-2)$ are the numbers of active-quark flavors (including $c$ and $b$ quarks as in Reference \cite{Gong:2012ah}) and light-quark flavors, respectively. In ${\rm SU}(3)$, the color factors are given by $T_F=\frac{1}{2}$, $C_F=\frac{4}{3}$, and $C_A=3$.

To calculate the $D$-dimension trace of fermion loops involving $\gamma_5$, under the scheme described in \cite{Korner:1991sx,z decay 4,z decay 22,Zheng:2017xgj}, we choose the same starting point ($Z$-vertex) to write down all the amplitudes without implementation of cyclicity.

\subsubsection{Real corrections}
The real corrections to $Z \to b\bar{b}[n]$+g ($n=^1S_0^{[8]},^3S_1^{[8]}$, and $^3P_J^{[8]}$) involve three 1$\to$3 processes,
\begin{eqnarray}
&Z& \to b\bar{b}[n]+g+g, \nonumber \\
&Z& \to b\bar{b}[n]+q+\bar{q}, \nonumber \\
&Z& \to b\bar{b}[n]+u_g+\bar{u}_g,
\label{CO real processes}
\end{eqnarray}
as illustrated in figure \ref{fig:real}. $``q"$ and $``u_g"$ represent the light quarks ($u,d,s$) and ghost particles, respectively. The phase-space integration of the three processes in equation (\ref{CO real processes}) will generate IR singularities, which can be isolated by slicing the phase space into different regions, namely the two-cutoff slicing strategy \cite{Harris:2001sx}. By introducing two small cut-off parameters ($\delta_s$ and $\delta_c$) to decompose the phase space into three parts, $\hat{\Gamma}_{\textrm{real}}$ can then be written as
\begin{eqnarray}
\hat{\Gamma}_{\textrm{Real}}&=&\hat{\Gamma}_{\textrm{S}}+\hat{\Gamma}_{\textrm{HC}}+\hat{\Gamma}_{\textrm{H}\overline{\textrm{C}}}.
\end{eqnarray}
$\hat{\Gamma}_{\textrm{S}}$ is the soft term arising only from the process of $Z \to b\bar{b}[n]+g+g$; $\hat{\Gamma}_{\textrm{HC}}$ denotes the hard-collinear term, which originates from all the three processes in equation (\ref{CO real processes}). The hard-noncollinear term $\hat{\Gamma}_{\textrm{H}\overline{\textrm{C}}}$ is IR finite and can be numerically computed by means of standard Monte-Carlo integration techniques. For $^1S_{0}^{[8]}$ and $^3S_{1}^{[8]}$, the combination with the IR singularities appearing in the virtual corrections would exactly cancel the soft divergences in the real processes; one can then obtain the finite $\hat{\Gamma}_{^1S_{0}^{[8]}}$ and $\hat{\Gamma}_{^3S_{1}^{[8]}}$. However, in the case of $^3P_{J}^{[8]}$, there remain residually soft singularities\footnote{Actually, in our NLO calculations, only the axial-vector part of $Z \to b\bar{b}[^3P_J^{[8]}]+g+g$ comprises the so-called residually soft singularities.}, which can only be removed by renormalizing $\langle \mathcal O ^{^3P_{J}^{[8]}}(^3S_{1}^{[8]})\rangle$, as in the case of $\langle \mathcal O ^{^3P_{J}^{[1]}}(^3S_{1}^{[8]})\rangle$ \cite{Jia:2014jfa}.

To derive $\hat{\Gamma}_{^3P_{J}^{[8]}}$, we first divide $\Gamma(^3P_{J}^{[8]})$ into two ingredients,
\begin{eqnarray}
\Gamma(^3P_{J}^{[8]})=\hat{\Gamma}_{^3P_{J}^{[8]}}\langle \mathcal O ^{^3P_{J}^{[8]}}(^3P_{J}^{[8]})\rangle+\hat{\Gamma}^{\textrm{LO}}_{^3S_{1}^{[8]}} \langle \mathcal O ^{^3P_{J}^{[8]}}(^3S_{1}^{[8]})\rangle^{\textrm{NLO}}.
\end{eqnarray}
Then one can obtain
\begin{eqnarray}
\hat{\Gamma}_{^3P_{J}^{[8]}}\langle \mathcal O ^{^3P_{J}^{[8]}}(^3P_{J}^{[8]})\rangle
&=&\Gamma(^3P_{J}^{[8]})-\hat{\Gamma}^{\textrm{LO}}_{^3S_{1}^{[8]}} \langle \mathcal O ^{^3P_{J}^{[8]}}(^3S_{1}^{[8]})\rangle^{\textrm{NLO}} \nonumber \\
&=&\left[\hat{\Gamma}_{\textrm{Born}}+\hat{\Gamma}_{\textrm{Virtual}}+(\hat{\Gamma}_{\textrm{S}}-\hat{\Gamma}_\textrm{S}^{\textrm{res}})+\hat{\Gamma}_\textrm{S}^{\textrm{res}}+\hat{\Gamma}_{\textrm{HC}}+\hat{\Gamma}_{\textrm{H}\overline{\textrm{C}}}+\hat{\Gamma}_{\textrm{NLO}^{*}}\right]\bigg|_{^3P_{J}^{[8]}} \nonumber \\
&\times & \langle \mathcal O ^{^3P_{J}^{[8]}}(^3P_{J}^{[8]})\rangle - \hat{\Gamma}^{\textrm{LO}}_{^3S_{1}^{[8]}} \langle \mathcal O ^{^3P_{J}^{[8]}}(^3S_{1}^{[8]})\rangle^{\textrm{NLO}},
\label{3pj8eq1}
\end{eqnarray} 
where $\hat{\Gamma}_\textrm{S}^{\textrm{res}}$ denotes the residually-soft-singularities involved terms in $\hat{\Gamma}_\textrm{S}$.

The soft singularities in $\hat{\Gamma}_\textrm{S}^{\textrm{res}}$ and $\langle \mathcal O ^{^3P_{J}^{[8]}}(^3S_{1}^{[8]})\rangle^{\textrm{NLO}}$ would cancel with each other, subsequently providing us with the finite $\hat{\Gamma}_{^3P_{J}^{[8]}}$,
\begin{eqnarray}
\hat{\Gamma}_{^3P_{J}^{[8]}}=\hat{\Gamma}_{\textrm{Born}}+\hat{\Gamma}_{\textrm{Virtual}}+\hat{\Gamma}_{\textrm{S}}^{*}+\hat{\Gamma}_{\textrm{HC}}+\hat{\Gamma}_{\textrm{H}\overline{\textrm{C}}}+\hat{\Gamma}_{\textrm{NLO}^{*}},
\end{eqnarray}
with
\begin{eqnarray}
\hat{\Gamma}_\textrm{S}^{*}=\left[(\hat{\Gamma}_{\textrm{S}}-\hat{\Gamma}^{\textrm{rev}}_{\textrm{S}})\big|_{^3P_{J}^{[8]}}+\frac{\alpha_s}{3 \pi m_b^2}u_{\epsilon}\frac{N_c^2-4}{N_c}\hat{\Gamma}^{\textrm{LO}}_{^3S_{1}^{[8]}}\right].
\end{eqnarray}
$u_{\epsilon}$ is finite, and its analytical expression can be found in Reference \cite{Jia:2014jfa}.

In calculating $\hat{\Gamma}_{^3P_{J}^{[8]}}$, the relations $\langle\mathcal O^{\Upsilon}(^3P_2^{[8]})\rangle=5\langle\mathcal O^{\Upsilon}(^3P_0^{[8]})\rangle$ and $\langle\mathcal O^{\Upsilon}(^3P_1^{[8]})\rangle=3\langle\mathcal O^{\Upsilon}(^3P_0^{[8]})\rangle$ are adopted.

\begin{figure}[!h]
\begin{center}
\hspace{0cm}\includegraphics[width=0.328\textwidth]{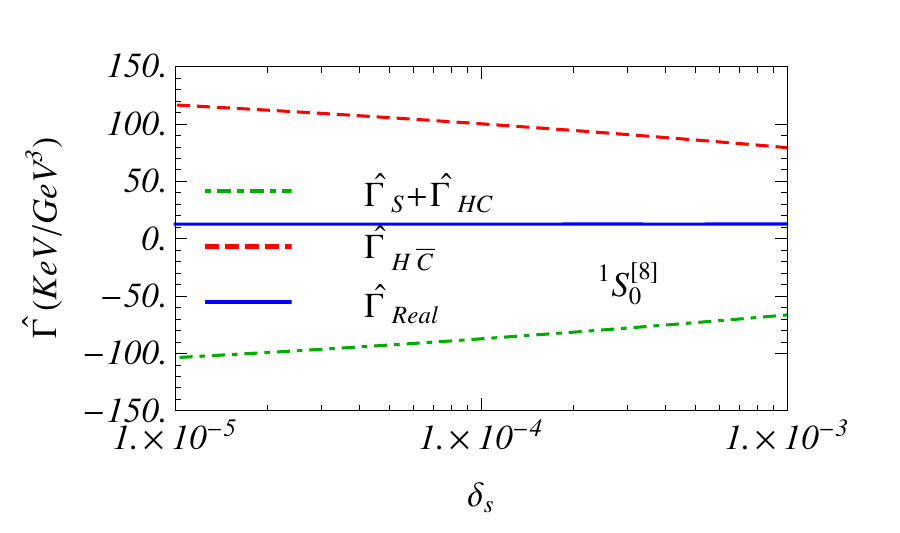}
\hspace{0cm}\includegraphics[width=0.328\textwidth]{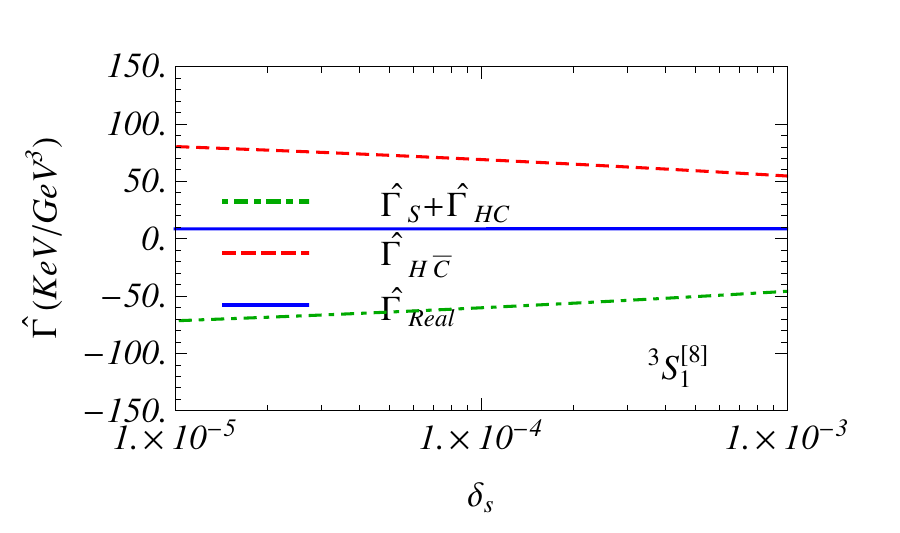}
\hspace{0cm}\includegraphics[width=0.328\textwidth]{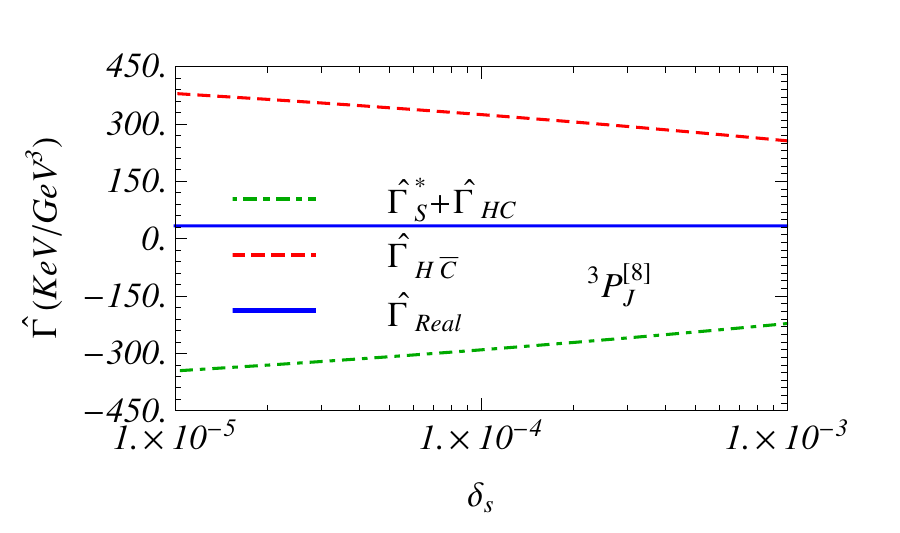}
\hspace{0cm}\includegraphics[width=0.328\textwidth]{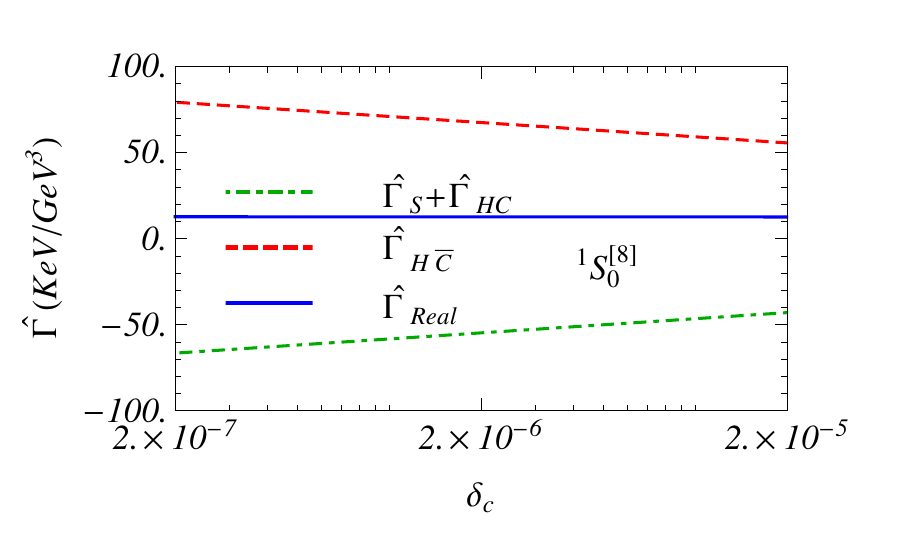}
\hspace{0cm}\includegraphics[width=0.328\textwidth]{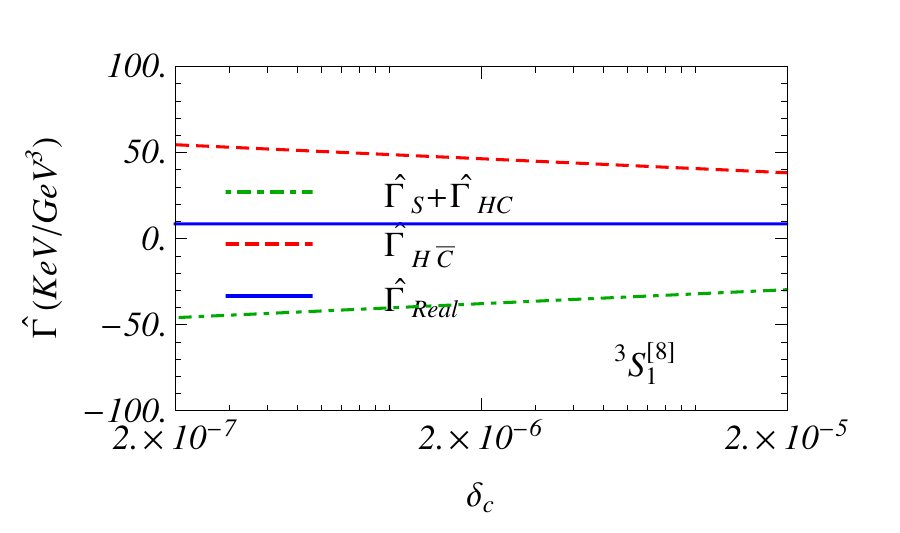}
\hspace{0cm}\includegraphics[width=0.328\textwidth]{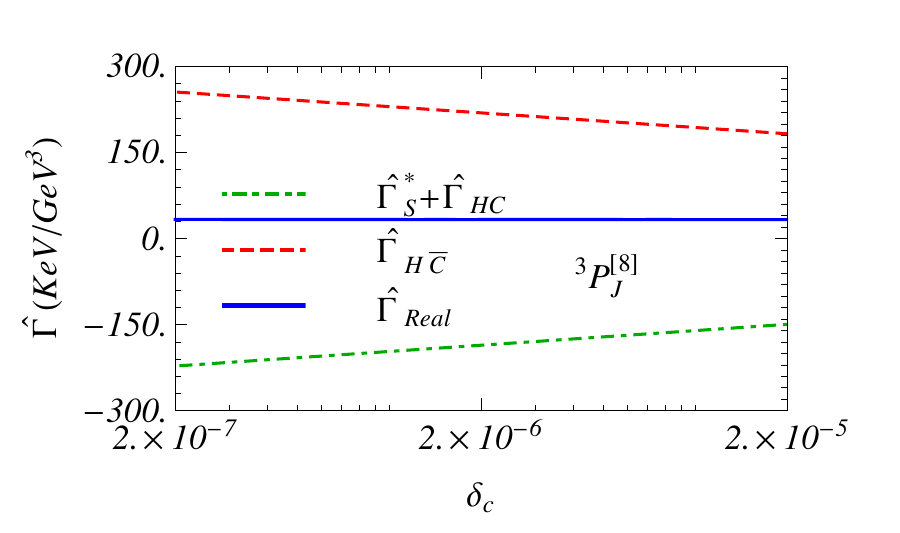}
\caption{\label{fig:cut independence}
Verifications of the independence on the soft ($\delta_s$) and collinear ($\delta_c$) cut-off parameters.}
\end{center}
\end{figure}

In our calculations, we use our $Mathematica$-$Fortran$ package with the implementation of FeynArts \cite{Hahn:2000kx}, FeynCalc \cite{Mertig:1990an}, FIRE \cite{Smirnov:2008iw}, and Apart \cite{Feng:2012iq}. This package has been employed to preform QCD corrections to several heavy-quarkonium related processes \cite{z decay 33,Sun:2017wxk,Sun:2018rgx}. Simultaneously, we apply another independent package, Feynman Diagram Calculation ($FDC$) \cite{Wang:2004du}, to compute all involved processes, and acquire the same numerical results. Independence on the cut-off parameters has been scrutinized, as is shown in figure \ref{fig:cut independence}\footnote{The process $Z \to q+\bar{q}+g^{*}; g^{*} \to b\bar{b}[^3S_1^{[8]}]$, which is divergence free and therefore does not depend on $\delta_{s,c}$, contributes significantly to $\hat{\Gamma}_{\textrm{H}\overline{\textrm{C}}}({^3S_1^{[8]}})$. In order to markedly demonstrate the verification of the independence on $\delta_{s,c}$, the $\hat{\Gamma}_{\textrm{H}\overline{\textrm{C}}}({^3S_1^{[8]}})$ in figure \ref{fig:cut independence} dose not include the contributions of this process.}. 

As a crosscheck, by taking the same input parameters, we have reproduced the NLO results of $\sigma ( e^+e^- \to c\bar{c}[^1S_0^{[8]},^3P_J^{[8]}]+g )$ in Reference \cite{Zhang:2009ym}. In addition, we have calculated the QCD corrections to $Z \to c\bar{c}[^3S_1^{[1]}]+\gamma$, whose one-loop and counterterm diagrams resemble that of $Z \to b\bar{b}[^3S_1^{[8]}]+g$, obtaining exactly the same $K$ factor given by Reference \cite{Wang:2013ywc}.

\section{Phenomenological results}\label{results}

\subsection{Input parameters}

The input parameters entering our calculations are taken as
\begin{eqnarray}
&&\alpha=1/128,~~~m_b=4.75~\textrm{GeV},~~~m_c=1.5~\textrm{GeV},\nonumber \\
&&m_{q/\bar{q}}=0~(q=u,d,s),~~~m_Z=91.1876~\textrm{GeV},\nonumber \\
&&\sin^{2}(\theta_W)=0.23116. \label{para}
\end{eqnarray}
As to the NRQCD LDMEs, we utilize three typical sets of these parameters \cite{LDMEs 1,LDMEs 2,LDMEs 3} to present the numerical results. In Refs. \cite{LDMEs 1} (Gong $et$ $al$.) and \cite{LDMEs 3} (Feng $et$ $al$.)\footnote{In our calculations, we choose $\mu_\Lambda=m_b$ and correspondingly take the LDMEs of table IV in Reference \cite{LDMEs 3}.}, the specific values of LDMEs have been provided; however, in Reference \cite{LDMEs 2} (Han $et$ $al$.), only the linear combination of $\langle \mathcal O ^{\Upsilon(nS)}(^1S_{0}^{[8]})\rangle$, $\langle \mathcal O ^{\Upsilon(nS)}(^3S_{1}^{[8]})\rangle$, and $\frac{\langle \mathcal O ^{\Upsilon(nS)}(^3P_{J}^{[8]})\rangle}{m_b^2}$ is given, i.e.,
\begin{eqnarray}
M_{0,r_0}^{\Upsilon(nS)}&=&\langle\mathcal{O} ^{\Upsilon(nS)}(^1S_0^{[8]})\rangle+r_0\frac{\langle\mathcal{O} ^{\Upsilon(nS)}(^3P_0^{[8]})\rangle}{m_b^2}, \nonumber \\
M_{1,r_1}^{\Upsilon(nS)}&=&\langle\mathcal{O} ^{\Upsilon(nS)}(^3S_1^{[8]})\rangle+r_1\frac{\langle\mathcal{O} ^{\Upsilon(nS)}(^3P_0^{[8]})\rangle}{m_b^2}.
\end{eqnarray}
With a view to applications of $M_0$ and $M_1$, we set $\langle\mathcal{O} ^{\Upsilon(nS)}(^1S_0^{[8]})\rangle$ to $\zeta M_{0,r_0}^{\Upsilon(nS)}$, and correspondingly $\frac{\langle\mathcal{O} ^{\Upsilon(nS)}(^3P_0^{[8]})\rangle}{m_b^2}=\frac{1-\zeta}{r_0}M_{0,r_0}^{\Upsilon(nS)}$, and vary $\zeta$ between 0 and 1 around the default value $1/2$.

The CS LDMEs, $\langle \mathcal O ^{\Upsilon(nS)}(^3S_1^{[1]})\rangle$ and $\langle \mathcal O ^{\chi_{bJ}(mP)}(^3P_J^{[1]})\rangle$, are related to the radial wave functions and their first derives at the origin ($n,m=1,2,3$) by the following formula:
\begin{eqnarray}
\langle \mathcal O^{\Upsilon(nS)}(^3S_1^{[1]}) \rangle&=&\frac{9}{2\pi}|R_{\Upsilon(nS)}(0)|^2, \nonumber  \\
\langle \mathcal O^{\chi_{bJ}(mP)}(^3P_J^{[1]}) \rangle&=&(2J+1)\frac{3}{4\pi}|R^{'}_{\chi_{bJ}(mP)}(0)|^2,
\end{eqnarray}
where $|R_{\Upsilon(nS)}(0)|^2$ and $|R^{'}_{\chi_{bJ}(mP)}(0)|^2$ read \cite{Eichten:1995ch}
\begin{eqnarray}
&&|R_{\Upsilon(1S)}(0)|^2=6.477~\textrm{GeV}^3,~~~|R_{\Upsilon(2S)}(0)|^2=3.234~\textrm{GeV}^3,\nonumber  \\ 
&&|R_{\Upsilon(3S)}(0)|^2=2.474~\textrm{GeV}^3, \nonumber \\ 
&&|R^{'}_{\chi_{bJ}(1P)}(0)|^2=1.417~\textrm{GeV}^5,~~~|R^{'}_{\chi_{bJ}(2P)}(0)|^2=1.653~\textrm{GeV}^5,\nonumber  \\ 
&&|R^{'}_{\chi_{bJ}(3P)}(0)|^2=1.794~\textrm{GeV}^5.
\end{eqnarray}

Branching ratios of $\chi_{bJ}(mP) \to \Upsilon(nS)$, $\Upsilon(nS) \to \chi_{bJ}(mP)$, $\Upsilon(3S) \to \Upsilon(2S)$, $\Upsilon(3S) \to \Upsilon(1S)$, and $\Upsilon(2S) \to \Upsilon(1S)$ can be found in Refs. \cite{LDMEs 1,LDMEs 2,LDMEs 3,Zyla:2020zbs}.

\subsection{SDCs}

\begin{figure}[!h]
\begin{center}
\hspace{0cm}\includegraphics[width=0.35\textwidth]{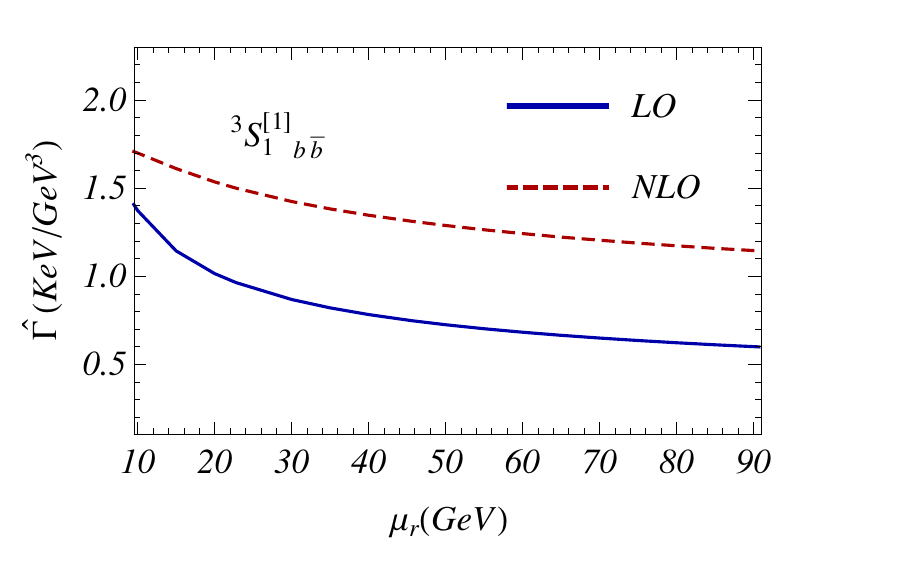}
\hspace{0cm}\includegraphics[width=0.35\textwidth]{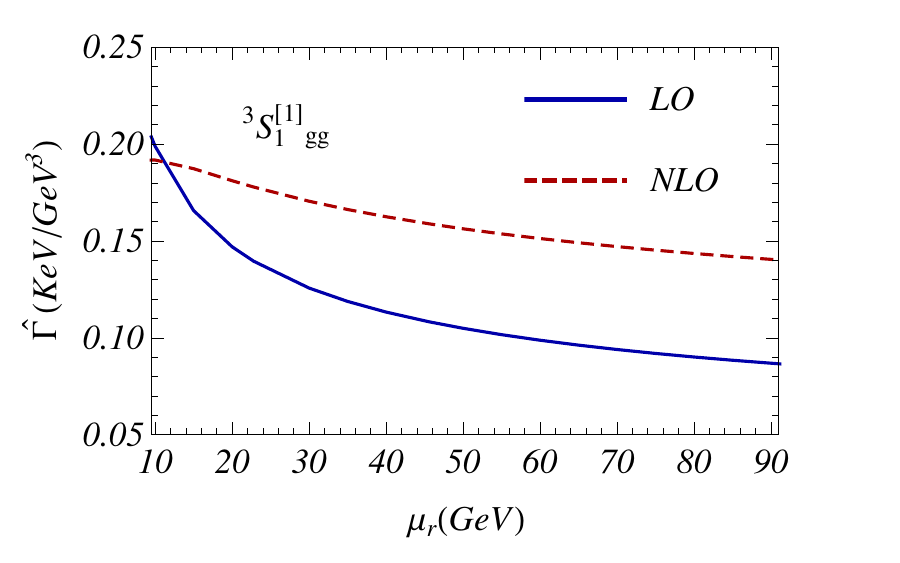}\\
\hspace{0cm}\includegraphics[width=0.35\textwidth]{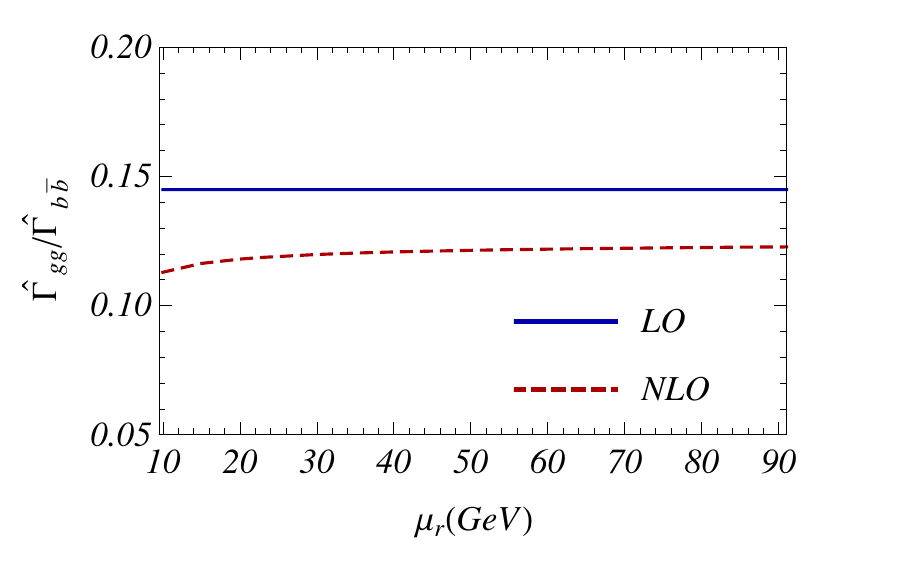}
\hspace{0cm}\includegraphics[width=0.35\textwidth]{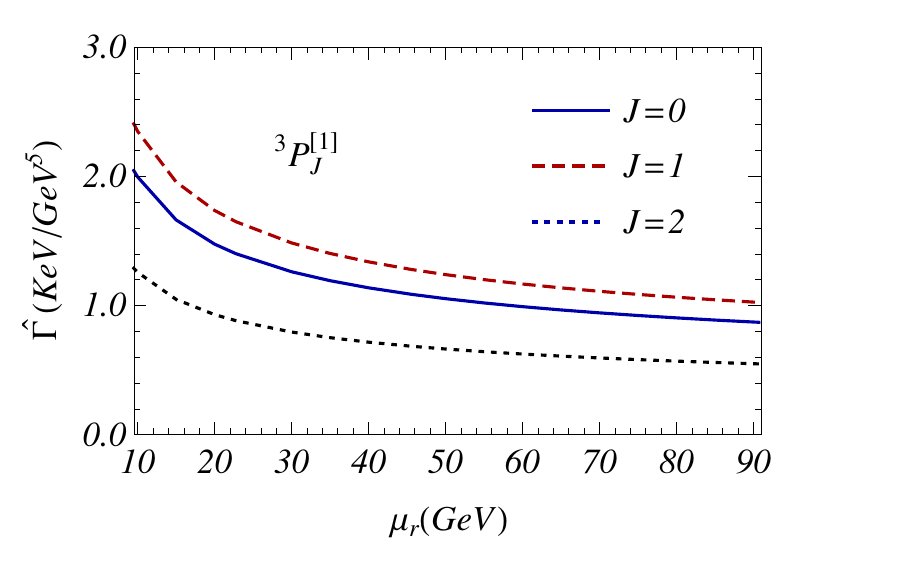}
\caption{\label{fig:CSsdc}
SDCs of the CS processes. The subscript $``gg"$ stands for $Z \to b\bar{b}[^3S_1^{[1]}]+g+g$ \cite{z decay 33}, and $``b\bar{b}"$ for $Z \to b\bar{b}[^3S_1^{[1]}]+b+\bar{b}$ \cite{z decay 4}, with $\hat{\Gamma}_{gg}/\hat{\Gamma}_{b\bar{b}}$ elucidating the ratio of these two channels. In deriving $\hat{\Gamma}_{^3P_J^{[1]}}$, we set $\langle \mathcal O^{\chi_{bJ}(mP)}(^3P_J^{[1]}) \rangle$ ($J=0,1,2$) equal to the same value of $\frac{3}{4\pi}|R^{'}_{\chi_{bJ}(mP)}(0)|^2$.}
\end{center}
\end{figure}

Before going further, we first take a look at the SDCs. To begin with, we summarize the CS SDCs in figure \ref{fig:CSsdc}. Inspecting the above two figures, the QCD corrections to the two $^3S_1^{[1]}$ processes in equation (\ref{cs processes}) are significant; moreover, the NLO results exhibit stronger stability than the LO ones under the renormalization scale variations. The third diagram tells us that the ${^3S_1^{[1]}}_{b\bar{b}}$ process plays a leading role in the $^3S_1^{[1]}$ predictions; however, ${^3S_1^{[1]}}_{gg}$ can also supply indispensable contributions, accounting for about $12\%$ of ${^3S_1^{[1]}}_{b\bar{b}}$ at the NLO accuracy. As illustrated in the last figure, $\hat{\Gamma}_{^3P_J^{[1]}}$ appears to be comparable in magnitude to $\hat{\Gamma}_{^3S_1^{[1]}}$; however, further multiplying by $\frac{3}{4\pi}|R^{'}_{\chi_{bJ}(mP)}(0)|^2$ and $\mathcal{B}_{\chi_{bJ} \to \Upsilon+\gamma}$ would eventually result in an insignificant $\chi_b$ feed-down contribution to $\Upsilon$ productions. That is to say, the $^3S_1^{[1]}$ processes are the ones which dominate the CS predictions.

\begin{figure}[!h]
\begin{center}
\hspace{0cm}\includegraphics[width=0.35\textwidth]{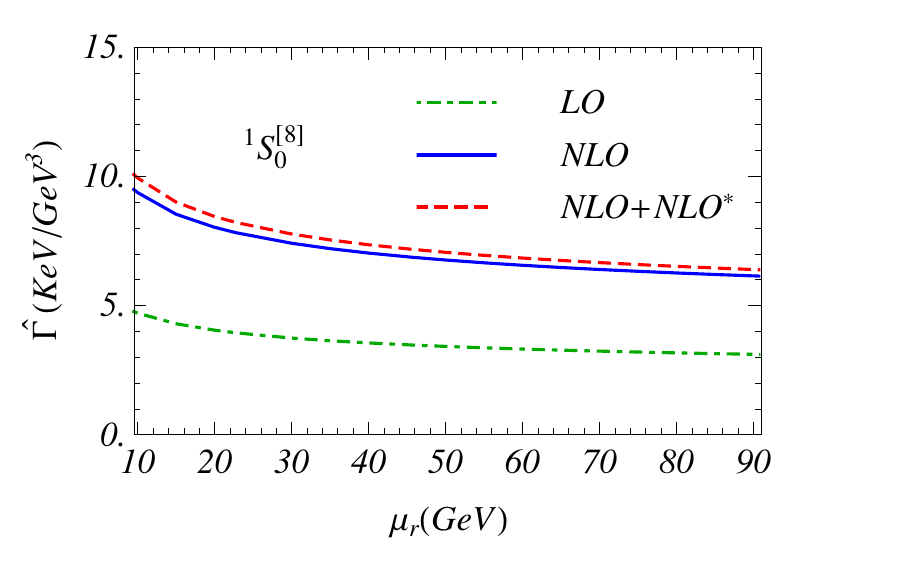}
\hspace{0cm}\includegraphics[width=0.35\textwidth]{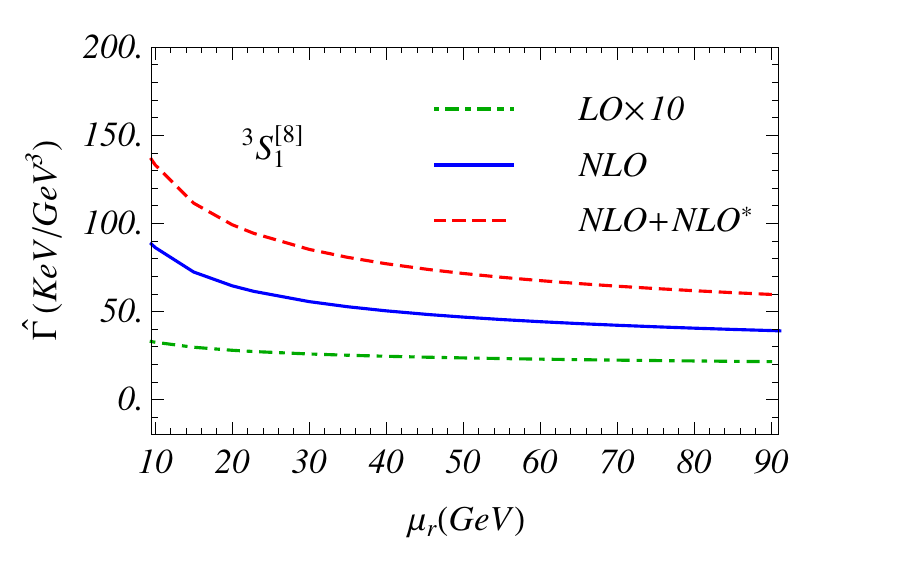}\\
\hspace{0cm}\includegraphics[width=0.35\textwidth]{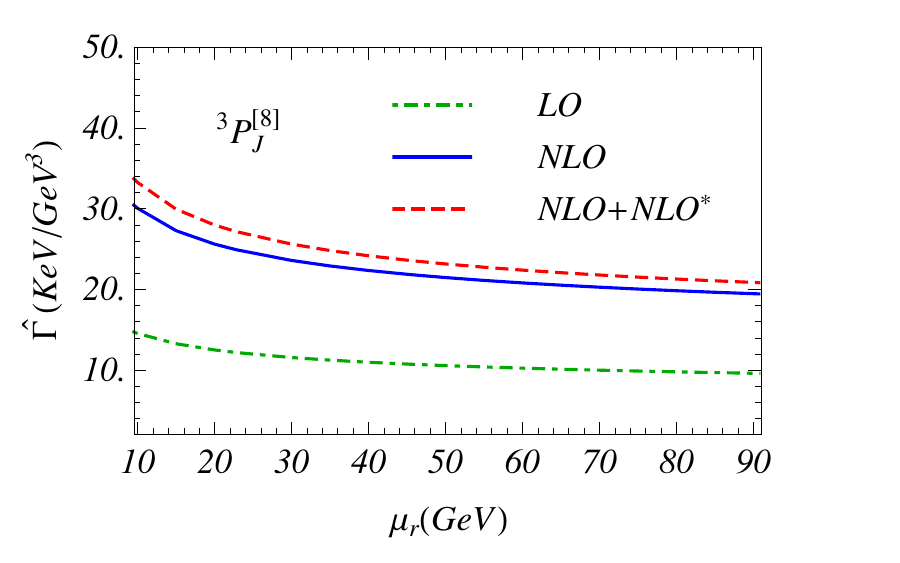}
\hspace{0cm}\includegraphics[width=0.35\textwidth]{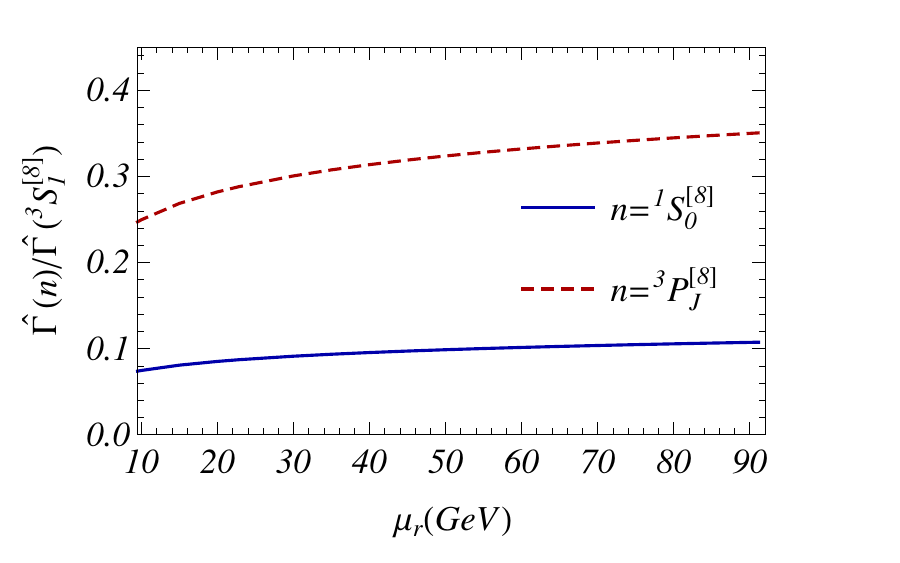}
\caption{\label{fig:COsdc}
SDCs of the CO processes.}
\end{center}
\end{figure}

The CO SDCs with respect to $\mu_r$ are drawn in figure \ref{fig:COsdc}. One can observe that the NLO QCD corrections to $Z \to b\bar{b}[^1S_0^{[8]},^3P_J^{[8]}]+X$ could enhance their LO results by about a steady factor of 2. The $\textrm{NLO}^{*}$ contributions given by $Z \to b\bar{b}[^1S_0^{[8]},^3P_J^{[8]}]+Q+\bar{Q}$ ($Q=c,b$) are slight. With regard to the $^3S_1^{[8]}$ state, the high-order terms in $\alpha_s$ can magnify its LO results to a extremely large extent, by about 25-30 times, which can be mainly ascribed to the gluon-fragmentation diagrams that first appear at the NLO level. By the same token, the $^3S_1^{[8]}$ $\textrm{NLO}^{*}$ processes further enlarge its NLO results by about $50\%$. The rightmost figure in the second column suggests that the SDC of $^3P_J^{[8]}$ is comparable to that of $^3S_1^{[8]}$; to be specific, the ratio of $\hat{\Gamma}_{^3P_J^{[8]}}$ to $\hat{\Gamma}_{^3S_1^{[8]}}$ can reach up to about $30\%$, increasing towards higher $\mu_r$. Although $\hat{\Gamma}_{^1S_0^{[8]}}$ is only about $10\%$ in magnitude of $\hat{\Gamma}_{^3S_1^{[8]}}$, the huge value of $\langle \mathcal{O}^{\Upsilon}(^1S_0^{[8]})\rangle$ \cite{LDMEs 1,LDMEs 2,LDMEs 3} would compensate for its smallness and provide a even larger contribution than that of $^3S_1^{[8]}$. Therefore, the existing theoretical studies, which just concentrate on the process of $Z \to q+\bar{q}+g^{*};g^{*} \to b\bar{b}[^3S_1^{[8]}]$, are indeed far insufficient to provide us with a thorough CO result.

\subsection{Decay widths}

\begin{table*}[htb]
\begin{center}
\caption{Decay widths of $Z \to \Upsilon(3S)+X$ (in units of KeV). The subscripts $``\textrm{Dr}"$ and $``\textrm{Fd}"$ refer to the direct-production processes and feed-down effects, respectively. $m_b=4.75$ GeV and $\mu_r=m_Z/2$.}
\label{tab: Upsilon3S}
\begin{tabular}{lccccc|cccccc}
\hline\hline
$~~~$ & $\Gamma_{\textrm{Dr}_{\textrm{CS}}}$ & $\Gamma_{\textrm{Dr}_{\textrm{CO}}}$ & $\Gamma_{\textrm{Fd}}$ & $\Gamma_{\textrm{Total}}$ & $\mathcal{B}(\times 10^{-6})$ & $\mathcal{B}_{\textrm{exp}}(\times 10^{-6})$ \cite{z decay 1}\\ \hline
Gong & $5.21$ & $2.01$ & $0.08$ & $7.30$ & $2.93$\\
Han & $5.21$ & $0.94$ & $0.30$ & $6.45$ & $2.59$ & $<94$\\
Feng & $5.21$ & $1.11$ & $0.37$ & $6.68$ & $2.68$\\ \hline \hline
\end{tabular}
\end{center}
\end{table*}

\begin{table*}[htb]
\begin{center}
\caption{Decay widths of $Z \to \Upsilon(2S)+X$ (in units of KeV). The subscripts $``\textrm{Dr}"$ and $``\textrm{Fd}"$ refer to the direct-production processes and feed-down effects, respectively. $m_b=4.75$ GeV and $\mu_r=m_Z/2$.}
\label{tab: Upsilon2S}
\begin{tabular}{lccccc|cccccc}
\hline\hline
$~~~$ & $\Gamma_{\textrm{Dr}_{\textrm{CS}}}$ & $\Gamma_{\textrm{Dr}_{\textrm{CO}}}$ & $\Gamma_{\textrm{Fd}}$ & $\Gamma_{\textrm{Total}}$ & $\mathcal{B}(\times 10^{-6})$ & $\mathcal{B}_{\textrm{exp}}(\times 10^{-6})$  \cite{z decay 1}\\ \hline
Gong & $6.82$ & $0.34$ & $2.05$ & $9.21$ & $3.69$\\
Han & $6.82$ & $1.51$ & $1.52$ & $9.85$ & $3.95$ & $<139$\\
Feng & $6.82$ & $2.19$ & $1.69$ & $10.7$ & $4.29$\\ \hline \hline
\end{tabular}
\end{center}
\end{table*}

\begin{table*}[htb]
\begin{center}
\caption{Decay widths of $Z \to \Upsilon(1S)+X$ (in units of KeV). The subscripts $``\textrm{Dr}"$ and $``\textrm{Fd}"$ refer to the direct-production processes and feed-down effects, respectively. $m_b=4.75$ GeV and $\mu_r=m_Z/2$.}
\label{tab: Upsilon1S}
\begin{tabular}{lccccc|cccccc}
\hline\hline
$~~~$ & $\Gamma_{\textrm{Dr}_{\textrm{CS}}}$ & $\Gamma_{\textrm{Dr}_{\textrm{CO}}}$ & $\Gamma_{\textrm{Fd}}$ & $\Gamma_{\textrm{Total}}$ & $\mathcal{B}(\times 10^{-6})$ & $\mathcal{B}_{\textrm{exp}}(\times 10^{-6})$  \cite{z decay 1}\\ \hline
Gong & $13.7$ & $0.34$ & $4.88$ & $18.9$ & $7.57$\\
Han & $13.7$ & $2.47$ & $4.50$ & $20.7$ & $8.30$ & $<44$\\
Feng & $13.7$ & $1.06$ & $5.09$ & $19.9$ & $7.98$\\ \hline \hline
\end{tabular}
\end{center}
\end{table*}

\begin{table*}[htb]
\begin{center}
\caption{Branching ratios of $\mathcal{B}_{Z \to \Upsilon(1S+2S+3S)+X}(\times 10^{-6})$. $m_b=4.75$ GeV and $\mu_r=m_Z/2$.}
\label{tab: Upsilon123S}
\begin{tabular}{lccccc|cccccc}
\hline\hline
$\mathcal{B}_{\textrm{Han}}$ & $\mathcal{B}_{\textrm{Gong}}$ & $\mathcal{B}_{\textrm{Feng}}$ & $\mathcal{B}_{\textrm{exp}}(\times 10^{-6})$ \cite{z decay 1}\\ \hline
$14.8$ & $14.2$ & $14.9$ & $100 \pm 40 \pm 22$\\ \hline \hline
\end{tabular}
\end{center}
\end{table*}

\begin{figure}[!h]
\begin{center}
\hspace{0cm}\includegraphics[width=0.328\textwidth]{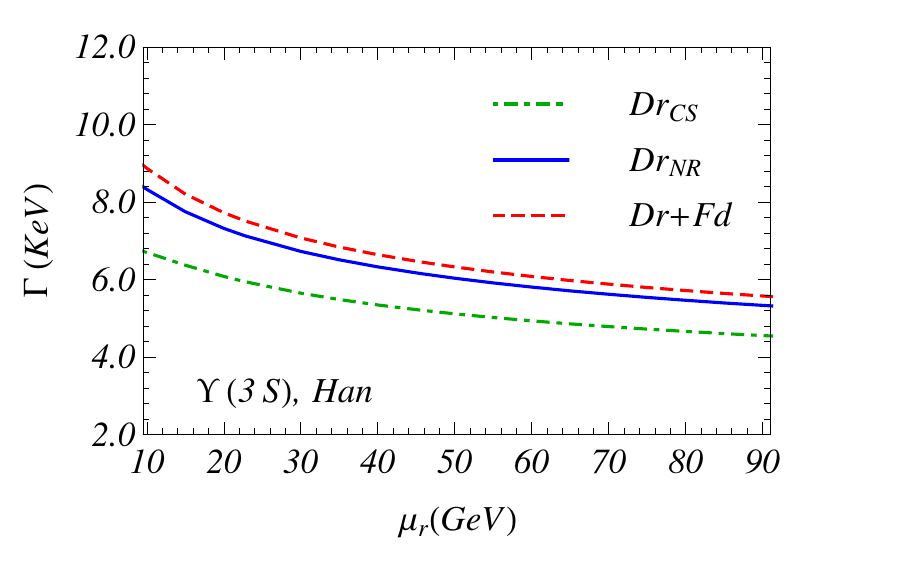}
\hspace{0cm}\includegraphics[width=0.328\textwidth]{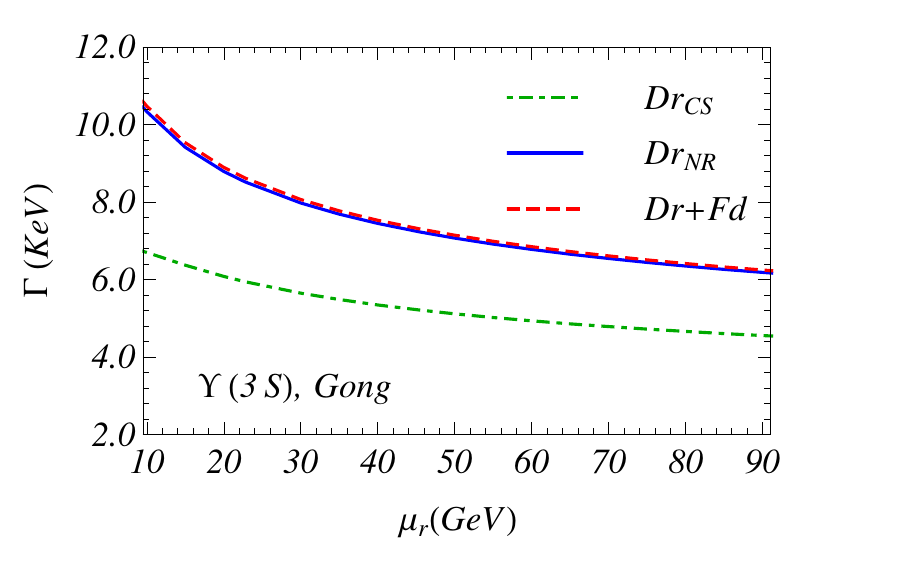}
\hspace{0cm}\includegraphics[width=0.328\textwidth]{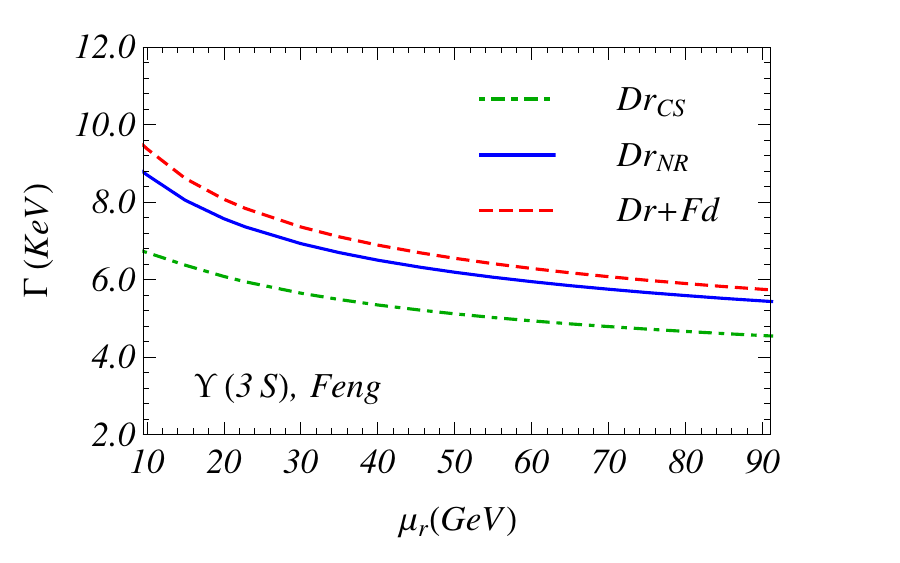}
\hspace{0cm}\includegraphics[width=0.328\textwidth]{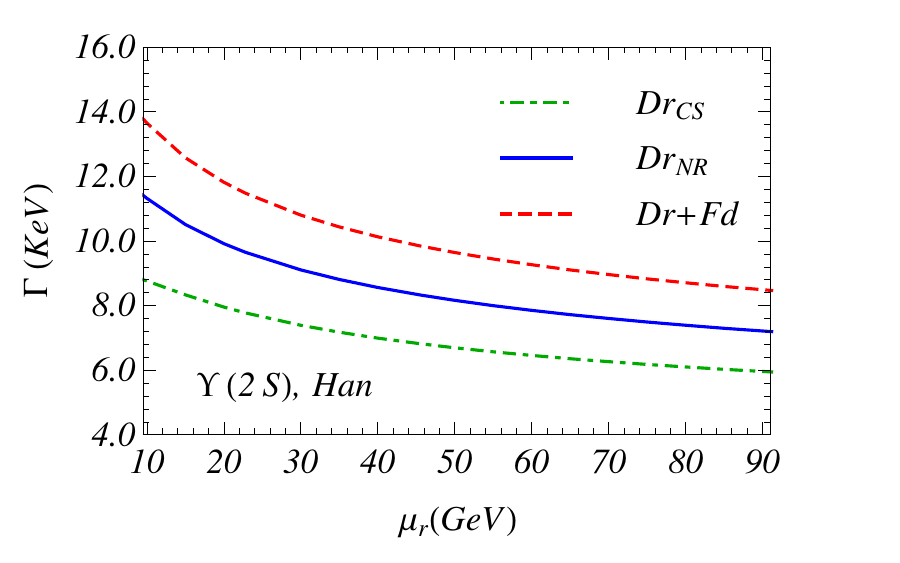}
\hspace{0cm}\includegraphics[width=0.328\textwidth]{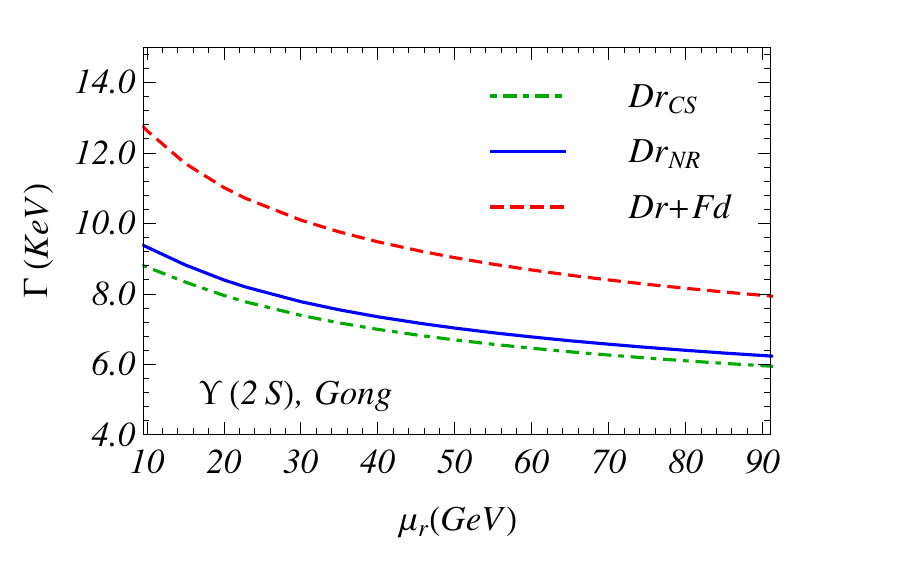}
\hspace{0cm}\includegraphics[width=0.328\textwidth]{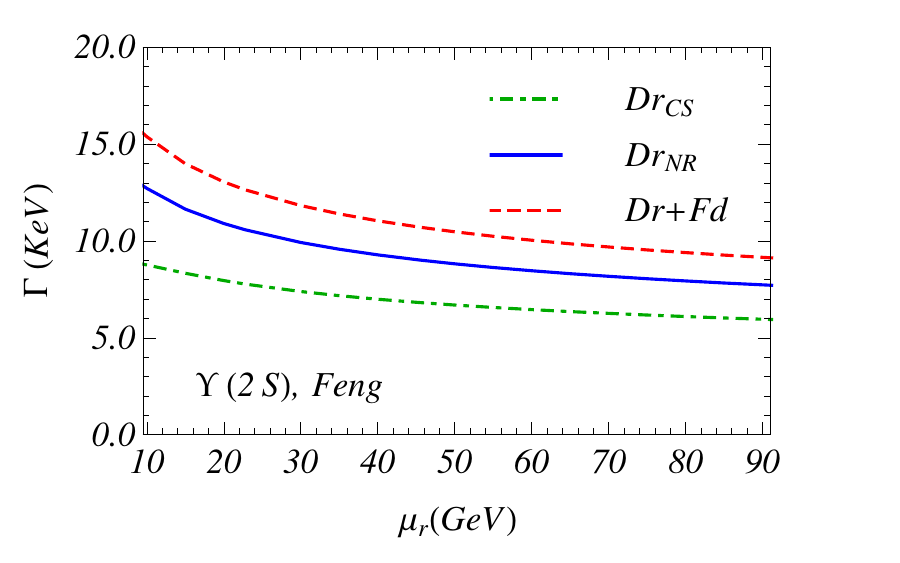}
\hspace{0cm}\includegraphics[width=0.328\textwidth]{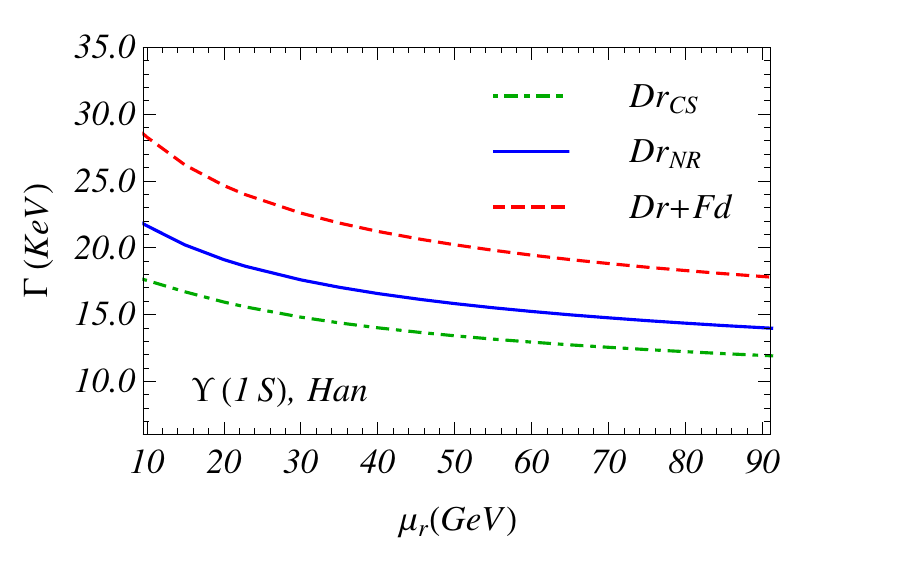}
\hspace{0cm}\includegraphics[width=0.328\textwidth]{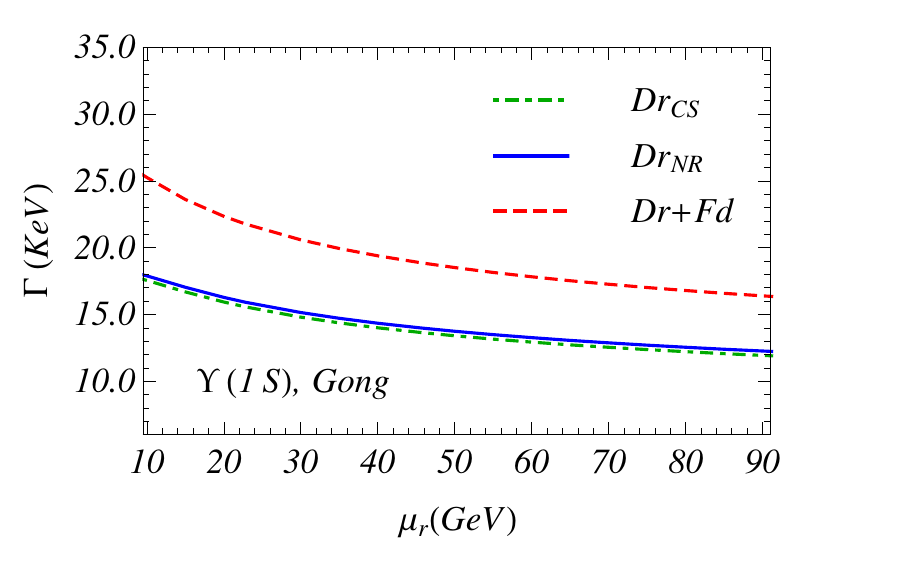}
\hspace{0cm}\includegraphics[width=0.328\textwidth]{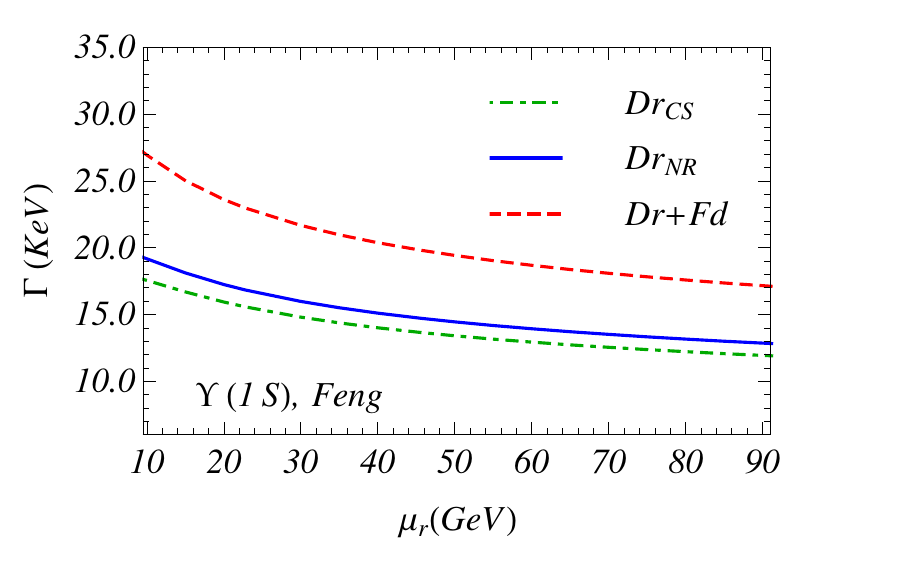}
\caption{\label{fig:decay width}
Decay widths of $Z \to \Upsilon(nS)+X$ as a function of $\mu_r$, obtained by the LDMEs of Refs. \cite{LDMEs 1,LDMEs 2,LDMEs 3}. The subscript $``\textrm{NR}"$ means the sum of the CS and CO contributions. $``\textrm{Dr}"$ and $``\textrm{Fd}"$ refer to the direct-production processes and feed-down effects, respectively. $m_b=4.75$ GeV and $\mu_r=m_Z/2$.}
\end{center}
\end{figure}

The NRQCD predictions of $\Gamma_{Z \to \Upsilon(nS)+X}$ are presented in tables. \ref{tab: Upsilon3S}, \ref{tab: Upsilon2S}, and \ref{tab: Upsilon1S}, and in figure \ref{fig:decay width}. We find
\begin{itemize}
\item[1)]
The CO contributions exhibit crucial influence on the theoretical predictions, the amount of which relying on the LDMEs choices. Taking the LDMEs of Refs. \cite{LDMEs 2} and \cite{LDMEs 3}, inclusion of the CO processes can increase the CS predictions ($\Gamma_{\textrm{Dr}_{\textrm{CS}}}$) by about $20-30\%$. When the LDMEs of Reference \cite{LDMEs 1} is adopted, $\Gamma_{\textrm{Dr}_{\textrm{CO}}}$ can magnify $\Gamma^{3S}_{\textrm{Dr}_{\textrm{CS}}}$ by about $40\%$; however, for $\Upsilon(2S)$ and $\Upsilon(1S)$ production, this kind of enhancements are mild.
\item[2)]
The feed-down effects via higher excited states are found to be considerable; for instance, the $\Upsilon(2,3S)$ and $\chi_b(1,2,3P)$ feed-down contributions can enlarge the $\Upsilon(1S)$ production by about $40\%$.
\end{itemize}
Taken together, our newly-calculated CO and feed-down contributions would introduce a $30-45\%$ enhancement to the existing theoretical results, manifesting the necessities of our calculations. By comparing with the L3 measurements, one can observe that the newest NRQCD predictions appear to be still at variance with the L3 data. To be specific, the predicted $\mathcal{B}_{Z \to \Upsilon(1S)+X}$ is about 5 times smaller in magnitude than the experimental upper limit, and this value can rise up to about 30 for $\Upsilon(2S,3S)$; in the case of $\mathcal{B}_{Z \to \Upsilon(1S+2S+3S)+X}$, the central value of the L3 measurement is about 7 times bigger in amount than our predictions. It is worth noting that the L3 Collaboration used only 6 reconstructed events to fit $\mathcal{B}_{Z \to \Upsilon(1S+2S+3S)+X}$, which is responsible, in part, for its large measuring uncertainties; moreover, the definite values of $\mathcal{B}_{Z \to \Upsilon(1,2,3S)+X}$, rather than the upper limits, have not yet been measured. In this sense, perhaps it is still premature to draw a decisive conclusion concerning the consistency or inconsistency of the NRQCD predictions with the L3 data, requiring future measurements at a collider with much higher luminosity, such as LHC or some planned $Z$ factories.

At last, we analyze the uncertainties of the predictions due to the choices of the renormalization scale $\mu_r$, the CO LDMEs, and the $b$-quark mass $m_b$.
\begin{itemize}
\item[(i)]
for $\mathcal{B}_{Z \to \Upsilon(3S)+X}(\times 10 ^{-6})$
\begin{eqnarray}
\mathcal{B}_{\textrm{Han}}&=&2.59^{+0.42+0.08+0.20}_{-0.36-0.08-0.18}, \nonumber \\
\mathcal{B}_{\textrm{Gong}}&=&2.93^{+0.53+0.09+0.24}_{-0.43-0.09-0.21}, \nonumber \\
\mathcal{B}_{\textrm{Feng}}&=&2.68^{+0.46+0.15+0.21}_{-0.38-0.17-0.20};
\end{eqnarray}
\item[(ii)]
for $\mathcal{B}_{Z \to \Upsilon(2S)+X}(\times 10 ^{-6})$
\begin{eqnarray}
\mathcal{B}_{\textrm{Han}}&=&3.95^{+0.66+0.15+0.30}_{-0.55-0.15-0.28}, \nonumber \\
\mathcal{B}_{\textrm{Gong}}&=&3.69^{+0.60+0.43+0.29}_{-0.51-0.43-0.26}, \nonumber \\
\mathcal{B}_{\textrm{Feng}}&=&4.29^{+0.78+0.29+0.35}_{-0.63-0.29-0.32};
\end{eqnarray}
\item[(iii)]
for $\mathcal{B}_{Z \to \Upsilon(1S)+X}(\times 10 ^{-6})$
\begin{eqnarray}
\mathcal{B}_{\textrm{Han}}&=&8.30^{+1.34+0.34+0.63}_{-1.14-0.34-0.57}, \nonumber \\
\mathcal{B}_{\textrm{Gong}}&=&7.57^{+1.16+0.30+0.58}_{-1.01-0.30-0.53}, \nonumber \\
\mathcal{B}_{\textrm{Feng}}&=&7.98^{+1.27+0.39+0.61}_{-1.08-0.39-0.57};
\end{eqnarray}
\item[(iv)]
for $\mathcal{B}_{Z \to \Upsilon(1S+2S+3S)+X}(\times 10 ^{-6})$
\begin{eqnarray}
\mathcal{B}_{\textrm{Han}}&=&14.8^{+2.42+0.56+1.14}_{-2.04-0.56-1.03}, \nonumber \\
\mathcal{B}_{\textrm{Gong}}&=&14.2^{+2.29+0.82+1.10}_{-1.95-0.82-1.00}, \nonumber \\
\mathcal{B}_{\textrm{Feng}}&=&14.9^{+2.50+0.82+1.17}_{-2.10-0.85-1.07},
\end{eqnarray}
\end{itemize}
where the three columns of uncertainties are caused by varying $\mu_r$ in $[2m_b,m_Z]$ around the central value of $m_Z/2$, varying the CO LDMEs from upper limit to lower limit around their central values \cite{LDMEs 1,LDMEs 2,LDMEs 3}, and varying $m_b$ in $[4.65,4.85]$ GeV around $4.75$ GeV, respectively. The above results indicate that the largest uncertainty lies in the ambiguities of the renormalization scale.
\section{Summary}\label{sum}

In order to tackle the great discrepancies between the theoretical results and the L3 data of $\Gamma_{Z \to \Upsilon+X}$, we in this manuscript revisit the inclusive $\Upsilon$ production in $Z$ decay by including the complete evaluations of the CO contributions at the QCD NLO accuracy for the first time. Our results show that the newly-calculated QCD corrections to the CO LO processes are significant, subsequently resulting in a large enhancement to the existing predictions given by the CS model. Besides, the feed-down contributions are considered exhaustively, and found to considerably enhance the decay width. Taking into account all theses contributions, the discrepancies between theory and L3 data remain conspicuous.

\acknowledgments
This work is supported in part by the Natural Science Foundation of China under the Grant No. 11705034, No. 12065006, and No. 11965006, and by the Project of GuiZhou Provincial Department of Science and Technology under Grant No. QKHJC[2019]1160 and No. QKHJC[2020]1Y035. \\

\providecommand{\href}[2]{#2}\begingroup\raggedright

\end{document}